\documentclass[twocolumn,prl,showpacs,floatfix,superscriptaddress]{revtex4}

\usepackage{amsmath,amssymb}
\usepackage{braket}
\usepackage{epsfig}
\usepackage[usenames,dvipsnames]{color}
\usepackage{hyperref}
\hypersetup{dvips,
  pdfauthor={Simone Chiesa, Christopher N. Varney, Marcos Rigol, and
    Richard T. Scalettar},
  pdftitle={Magnetism and pairing of two-dimensional trapped fermions},
  letterpaper,
  colorlinks=true,
  linkcolor=blue,
  citecolor=blue,
  pdfpagemode=UseNone
}

\begin{document}

\title{Magnetism and pairing of two-dimensional trapped fermions}

\author{Simone Chiesa}
\affiliation{Department of Physics and Astronomy, University of
  Tennessee, Knoxville, TN 37996, USA}
\affiliation{Department of Physics, University of California, Davis,
  CA 95616, USA} 

\author{Christopher N. Varney}
\affiliation{Department of Physics, Georgetown University, Washington,
  DC 20057, USA} 
\affiliation{Joint Quantum Institute, University of Maryland, College
  Park, MD 20742, USA} 

\author{Marcos Rigol}
\affiliation{Department of Physics, Georgetown University, Washington,
  DC 20057, USA} 

\author{Richard T. Scalettar}
\affiliation{Department of Physics, University of California, Davis,
  CA 95616, USA} 

\begin{abstract}
  The emergence of local phases in a trapped two-component Fermi gas
  in an optical lattice is studied using quantum Monte Carlo
  simulations.  We treat temperatures that are comparable or lower
  than those presently achievable in experiments and large enough
  systems that both magnetic and paired phases can be detected by
  inspection of the behavior of suitable short-range correlations.  We
  use the latter to suggest the interaction strength and temperature
  range at which experimental observation of incipient magnetism and
  $d$-wave pairing are more likely 
  and evaluate the relation between entropy and temperature in
  two-dimensional confined fermionic systems.
\end{abstract}

\maketitle

Strong electron correlations in solids are believed to be at the
origin of a remarkable host of phenomena that range from anomalous
insulating states and magnetism to high temperature superconductivity
\cite{bednorz86}. The Hubbard Hamiltonian \cite{hubbard63}, a model
which provides a simplified picture of the band structure and electron
interactions, is thought to contain the necessary ingredients to
describe such spectacular diversity.  In the strongly interacting
limit the half-filled model describes a Mott insulator and, on a
two-dimensional (2D) square lattice, quantum Monte Carlo simulations
\cite{hirsch89,varney09} have convincingly established the existence
of antiferromagnetism at $T=0$.  This conjunction of insulating
behavior and long-range antiferromagnetic order accurately describes
the low temperature physics of the undoped parent compounds of the
cuprate superconductors.

Despite intensive analytical and computational work, what happens as
one dopes the antiferromagnet has remained controversial and many
important questions remain unanswered
\cite{scalapino06,anderson07,lee06}.  A promising new route to study
the model has recently emerged in the form of experiments with
fermionic gases in optical lattices.  The appeal of these experiments
lies in the high degree of control over the different parameters of
the system and the ensuing possibility of a close realization of a
``pure'' Hubbard Hamiltonian \cite{bloch08}, free of the complexity
characterizing solid state systems.  Ground breaking achievements
include loading an ideal quantum degenerate Fermi gas in a
three-dimensional (3D) lattice \cite{kohl05} and the realization of
the Mott metal-insulator transition in 3D
\cite{jordens08,schneider08}.

Optical lattice experiments require the presence of a confining
potential, usually harmonic, and are accordingly modeled by the
Hamiltonian
\begin{equation}
  H\! = \!-t \!\sum_{\langle {i j} \rangle, \sigma}\!\!\left( c^\dag_{j
    \sigma} c_{i \sigma}^{\phantom\dag} +\mathrm{H.c.}\right)\!
   + \sum_{i}\!\left( V_i n_{i}  + U n_{ i \uparrow} n_{i \downarrow}\right),
  \label{hubham}
\end{equation}
with $V_i\equiv Vr_i^2$. The curvature $V$ causes the
density to vary across the lattice, resulting in a situation at
odds with the homogeneous system that one would like to emulate.
This is a potential problem because numerical studies on related 
models \cite{white98} have 
indicated the existence of many competing phases with vastly different physical
properties.  Understanding the extent to which the delicate balance
between these phases is affected by the presence of the trap is,
therefore, of importance for assessing whether fermionic gases can be
taken as accurate simulators of homogeneous models. Thanks to
improvements in quantum Monte Carlo (QMC) codes
\cite{vollhardt93,georges96,hettler98,rubtsov05,varney09}, this and
other issues can now be quantitatively investigated at temperatures
that are comparable or below those of the latest optical lattice
experiments \cite{jordens08,schneider08}.

Here we report results from state of the art determinant QMC (DQMC)
simulations \cite{blankenbecler81} of the trapped 2D Fermi-Hubbard
Hamiltonian. We found clear signatures of
magnetic and pairing correlations at temperatures of the order of the
magnetic scale $J=4t^2/U$, a perhaps surprising observation considering 
that the corresponding temperature range in strongly correlated solid state 
systems is well above $T_c$. We note, however,
that the properties that we compute are local in character, and that
recent advances in optical lattice experiments \cite{Gemelke09,Bakr09,Sherson10}
have made possible the imaging of individual sites and short range
correlations around them. Hence, our results suggest that the purity
of optical lattices and the novel probes used in these experiments
could allow the observation of local pairing signatures at higher
temperatures than possible in a condensed matter environment.

\begin{figure}
\includegraphics[width=0.45\textwidth,angle=0]{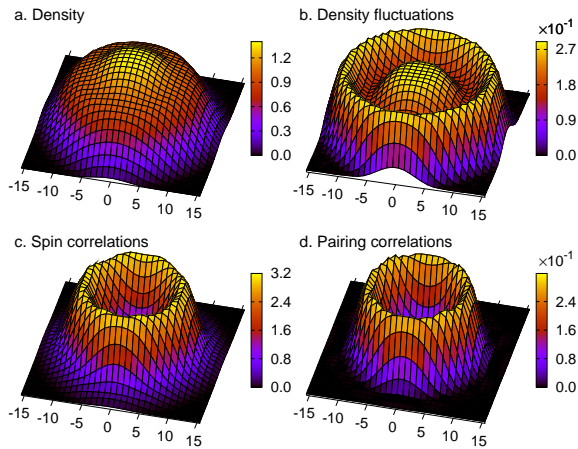}
\vspace{-0.2cm}
\caption{\label{Fig1} (a) Local density $n(i)$, (b) density
  fluctuations $\delta(i)$, (c) local staggered magnetization $M(i)$,
  and (d) $d$-wave pairing $\chi(i)$ are shown for $U =6t$, $T =
  0.31t$, $\mu_0 = 3.0t$, and $V = 0.04t$.  A Mott
  insulating domain is emerging in the density profile of
  panel (a), in the form of a half-filled ring 6-10 lattice spacings
  from the trap center.  The density fluctuations are minimized in
  this region. The staggered magnetization and the $d$-wave pairing,
  however, show a pronounced maximum in the Mott domain.}
\end{figure}

We analyze the properties of the trapped system in terms of the
spatial dependence of the local density $n(i) = \langle n_i\rangle$, 
the density fluctuations $\delta(i) = \langle n_{i}^2 \rangle - \langle
n_{i}^{\phantom 2} \rangle ^2$, and the spin correlations 
$C_{xy}(i)=4\langle S^z_{i+(x,y)} S^z_i\rangle$. Pairing is analogously
discussed by defining a local $d$-wave pair creation operator
${\Delta^d_i}^\dagger$
\begin{align}
  \begin{split}
    \label{fulld}
    \Delta^\dagger_{ij} &= \frac{1}{\sqrt{2}}
    \left(c^{\dagger}_{i\uparrow} c^{\dagger}_{j\downarrow} +
    c^{\dagger}_{j\uparrow} c^{\dagger}_{i\downarrow}\right)\\ 
    {\Delta^d_i}^\dagger &= \frac{1}{2} (\Delta^\dagger_{i,i+\hat{x}}
    + \Delta^\dagger_{i,i-\hat{x}} - \Delta^\dagger_{i,i+\hat{y}} -
    \Delta^\dagger_{i,i-\hat{y}})
  \end{split}
\end{align}
and a corresponding correlation function $P_{xy}(i)=\langle
\Delta^{d}_{i+(x,y)} \Delta^{d\dagger}_{i}\rangle$.  To isolate the
effect of the pairing vertex \cite{white89}, we focus on the connected
correlation function in which suitable products of single particle
propagators are subtracted from the expression in Eq.~\eqref{fulld}.

In Fig.~\ref{Fig1}, we examine the spatial dependence of different
properties at $U = 6\,t$, $V=0.04\,t$ and $T =0.31\, t$ for a system
of 560 fermions.  In this parameter regime, the average sign in DQMC
is $0.3$ and decreases exponentially as $T$ is lowered. $T=0.31\, t$
is therefore very close to the bound that the sign problem
\cite{loh90,troyer05} imposes on the lowest achievable
temperature. Despite this restriction, the emergence of a Mott
insulator is clearly visible and depicted in Fig.~\ref{Fig1}(a), which
shows a density plateau region of commensurate filling, $n(i)\simeq
1$, and in Fig.~\ref{Fig1}(b) through the formation of a pronounced
minimum in the density fluctuations. This domain is also distinguished
by enhanced antiferromagnetism [Fig.~\ref{Fig1}(c)] and $d$-wave
pairing [Fig.~\ref{Fig1}(d)]. These last two properties are
represented by the local order parameters
\begin{equation*}
  M(i)=\sum_{x,y} (-1)^{x_i+y_i+x+y}C_{x,y}(i),\quad \chi(i)=\sum_{x,y}
  P_{x,y}(i) ,
\end{equation*}
which are expected to diverge if long-range order develops around site
$i$.

Given that the temperature $T= 0.31\,t$ is of the order of the
antiferromagnetic exchange $J$, the sharp signal in $M(i)$ is clearly
due to the formation (and partial ordering) of local moments in the
Mott domain. The appearance of the peak in $\chi(i)$ in the same region
is however surprising - one would expect the peak to occur away from
the insulating phase. This shows that $M(i)$ or $\chi(i)$ can be
accurate indicators of the appearance of local phases only at
temperatures sufficiently low that the order parameter is dominated by
long-range contributions. As Fig.~\ref{Fig1}(d) demonstrates, this is
not the case in our simulations nor in current experiments.  However
we shall argue that the temperature and interaction dependence of spin
and pairing correlations, when examined at distances which are
sensitive to the appropriate energy scales, can provide compelling
evidence of incipient order \cite{supmat}.

\begin{figure}
 \centering
 \includegraphics[width=0.45\textwidth,angle=0]{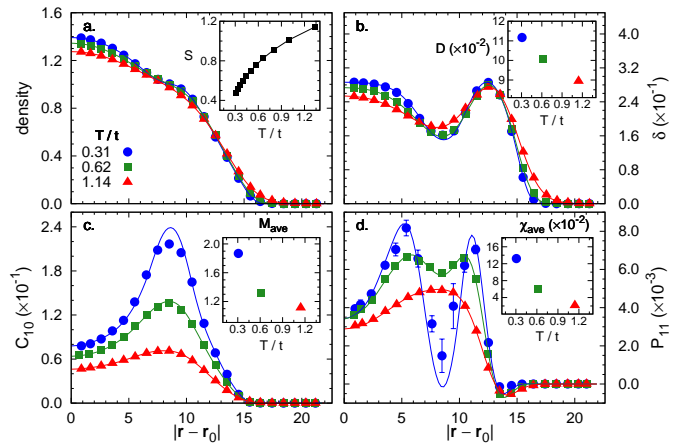}
\vspace{-0.2cm}
 \caption{\label{Fig2} Temperature dependence of the (a) density, (b)
   density fluctuations, (c) nearest-neighbor spin correlations
   $C_{10}(i)$, and (d) next-nearest-neighbor $d$-wave pairing
   $P_{11}(i)$ for $U=6t$. Lines are results within the local density
   approximation. The insets of (a) and (b) show the temperature
   dependence of the entropy and the double occupancy normalized to
   the number of particles, respectively, while the insets of (c) and
   (d) are the averages over the lattice of the local staggered
   magnetization and $d$-wave pairing.}
\end{figure}

Using this argument, we proceed to determine the temperature and
entropy scale that experimentalists need to achieve in trapped 2D
lattices in order to observe the onset of antiferromagnetic and
pairing correlations. Note that a related analysis has been recently
carried out for the entropy scale of the {\it half-filled} homogeneous model
\cite{paiva10}.

As the temperature is lowered from $T=1.14\,t$ to $T=0.31\,t$
($U=6\,t$), the density distribution [Fig.~\ref{Fig2}(a)] is slightly
compressed, the entropy per particle $S$ [shown in Fig.~\ref{Fig2}(a)
inset] decreases monotonically, and there is a marked increase in the
total double occupancy in the system $D$ [see inset in
Fig.~\ref{Fig2}(b)], which is generated by the increase of the double
occupancy at the trap center.  We determine the entropy, $S =
\beta(\langle E \rangle-F)$, by first computing the Helmholtz free
energy using constant-temperature coupling-constant integration over
an identical set of traps containing approximately $560$ fermions and
increasing $U$ \cite{supmat}.
The Mott insulating region $n(i)\simeq 1$ is characterized by a
deepening of the minimum in the density fluctuations $\delta(i)$
[Fig.~\ref{Fig2}(b)] and by a three-fold increase in the
nearest-neighbor (n.n.) spin-spin correlation $C_{10}(i)$
[Fig.~\ref{Fig2}(c)].  As the latter may be the easiest way to observe
the onset of antiferromagnetic correlations in experiments,
Fig.~\ref{Fig2}(c) makes clear that, although a weak signal may be
visible at the highest $T$, a clear maximum in the Mott domain will
only be apparent when $T<t$.

When analyzing pairing correlations it is important to realize that
$P_{00}(i)$ and $P_{10}(i)$ contain large contributions from
$C_{10}(i)$ and, consequently, carry little information on the actual
pairing amplitude in the vicinity of a given site. This is analogous
to the more familiar statement that local moment formation cannot be
taken as indication of short range magnetic order.  $P_{11}(i)$
[Fig.~\ref{Fig2}(d)] is therefore the shortest-distance correlation
that can be used to gain insights on the development of off-diagonal
order; it is characterized by two maxima around the Mott region,
sharply peaked at densities ($n=0.8$ and $n=1.2$) that lie in the
center of the superconducting dome of the cuprate phase diagrams
\cite{krockenberger08}.  The analogous correlation for pairs in an
extended $s$-wave state (see \cite{supmat}) is also significantly
enhanced in a broad ring around $n=0.6$. Although the local character
of $P_{11}$ and the high temperature do not allow a characterization
of the dominant low-temperature pairing symmetry, our results suggest
that $d$-wave and extended $s$-wave pairing may be detected as leading
pairing symmetries in spatially distinct regions, in agreement with
existing ground state calculations on related homogeneous models
\cite{rigol09,dagotto94}.

\begin{figure}
 \centering
 \includegraphics[width=0.45\textwidth,angle=0]{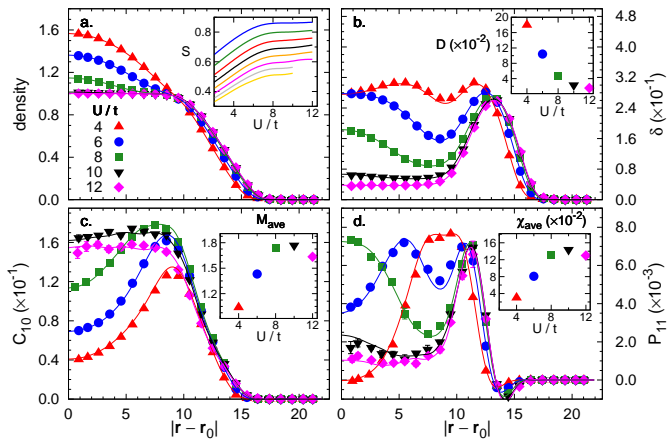}
\vspace{-0.2cm}
 \caption{ Same quantities of Fig.~\ref{Fig2} but as a function of
   interaction strength $U/t$ at constant $T=0.50t$. The inset of
   panel (a) shows the entropy for temperatures $T / t = 0.67, 0.57,
   0.50, 0.44, 0.40, 0.36, 0.33$, and $0.31$ (top to bottom).}
 \label{Fig3} 
\end{figure}

Another important piece of information that will help experimentalists
observe antiferromagnetism and pairing is the optimal value of the
ratio $U/t$ at which those correlations are expected to be maximal. 
In order to address this issue, we first investigate the interaction
dependence of the same physical properties represented in 
Fig.~\ref{Fig2} but at constant $T = 0.5\,t$ (Fig.~\ref{Fig3}),
and then show that our
conclusions are unaltered if it is $S$ to be held constant.

At intermediate coupling ($U=4\,t$), the smallness of the gap, the
curvature of the trapping potential, and thermal fluctuations make it
impossible for a flattening in the density profile to appear.  Note,
however, that signals of incipient insulating behavior are clear in
the structure of density fluctuations [Fig.~\ref{Fig3}(b)]. At large
interaction strength ($U=10\,t,12\,t$), the existence of a Mott
insulating domain can be directly inferred by inspection of the
density profile: the large central plateau comprising about 300 sites
is indicative of a phase with vanishing compressibility.
Nearest-neighbor spin correlations acquire their largest value at
$U=8\,t$. While the double occupancy is monotonically suppressed as
$U$ increases, the staggered magnetization shows a similar behavior to
$C_{10}(i)$, reaching a maximum when $U$ is of the order of the
bandwidth.  It is therefore clear that the optimal interaction
strength must be near the bandwidth and that future experiments should
focus their attention on this regime.  Similarly $P_{11}(i)$ exhibits
a sharp maximum at $n(i) = 0.8$ for $U \simeq 8\, t$ but contrary to
what is observed for $C_{10}(i)$, there is no weakening when going
from $U=8t$ to $U =12t$ and no dependence of the ``optimal'' density
for pairing on $U$. The latter is reminiscent of the properties of
cuprate superconductors where optimal doping is constant across the
entire family of these materials.

One might wonder what would happen in the experimentally relevant case of a 
(quasi-)adiabatic evolution, i.e., 
a system at constant entropy, as a function of $U$. In the inset of Fig.~3(a), 
we show the evolution of the entropy per particle as a function of the interaction 
strength for various fixed temperatures. As $U$ is increased at constant $S$, we find 
that the system cools down and goes, for instance, from $T=0.5t$ at $U=2t$ to 
$T=0.31t$ at $U=10t$, while $S$ is held fixed at 0.5. Since this cooling is mild beyond 
$U=8t$, isothermal and adiabatic evolution in this parameter regime are essentially 
equivalent. For $U<8t$, an adiabatic increase of $U$ is accompanied by progressively lower 
temperatures and this, in turn, implies that $U=8t$ remains the optimal interaction 
strength to observe strong-correlation phenomena.

We finally touch on the extent to which properties in a trap are a
faithful representation of those in a homogeneous system. To this aim
we use DQMC simulations on homogeneous $8\times8$ clusters to estimate
local properties as a function of chemical potential and plot the
corresponding local density approximation (LDA) results in
Figs.~\ref{Fig2} and \ref{Fig3} as continuous lines. The agreement
with the ab-initio results is excellent with the sole exception of the
half-filled region at $T=0.31\,t$ [Fig.~\ref{Fig2}(c)]. This
discrepancy is a characteristic failure of the LDA when applied to the
interface between coexisting phases \cite{rigol03Fa}. 

\begin{figure}[!htb]
 \centering
 \includegraphics[width=0.45\textwidth,angle=0]{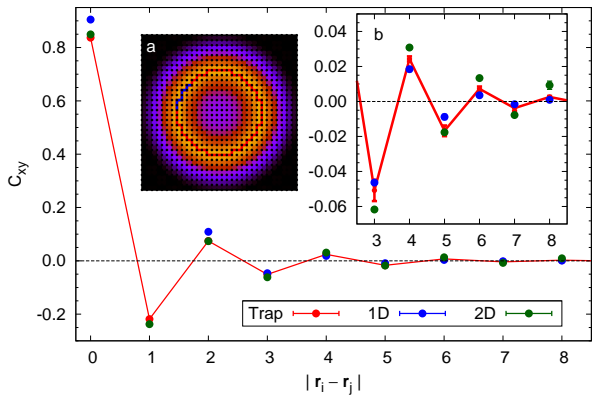}
\vspace{-0.2cm}
\caption{\label{Fig4} Spin correlations $C_{xy}(i)$ [for fixed site
    $i$ and variable separation $(x,y)$] along the path illustrated in
  inset (a) around the trapped center.  Inset (b) is an enlarged view
  of the longer correlations.  $C_{xy}(i)$ exhibits the alternating
  sign characteristic of antiferromagnetism.  The correlations
  decrease rapidly, but nevertheless extend out to several lattice
  spacings. The spin correlations are also shown in the LDA
  approximation for $d = 1$ and $d = 2$.}
\end{figure}

A more dramatic failure of the LDA is expected when examining longer
range correlations. Since in the fermionic Hubbard model the
development of a certain type of order corresponds with a rather
narrow density interval, different phases in a trap appear with a
narrow-ring-like topology.  If one disregards interfacial effects,
these regions can be thought as one-dimensional homogeneous strips
where the long range character of correlations is determined by the
strip width.  An example of such a situation is given by the
development of long-range magnetic order in the half-filled annulus
found in our simulations at $U=6t$.  In particular, Fig.~\ref{Fig4}
shows that 2D correlations, computed on a half-filled homogeneous
cluster at the same temperature, represent a good approximation to the
correlations in a trap only at short distances, overestimating the
correct long-range value. Analogous results for a 1D chain fail at all
distances and, in particular, decay too quickly at large separation.
This behavior is generic and expected for the superfluid regions as
well.  When the system size increases, these quasi-1D regions grow
both in width and length and their correlations ultimately cross over
to a pure 2D character. This dimensionality effect must be taken into
account when using finite size systems to infer the critical behavior.

In summary, we have addressed finite temperature properties of inhomogeneous
Fermi-Hubbard systems using an ab-initio approach. Our findings of
enhanced antiferromagnetic and pairing correlations just below the
temperature scale $T \sim t$ thus open important perspectives for
current experiments with ultracold fermions in optical lattices.
While the lowest temperatures reported here are well above the
$d$-wave coherence temperature, the enhanced signal in local
properties is a promising sign: these results and the purity of the
experimental optical lattice suggest that, in contrast to solid state
systems, temperatures of the order of the hopping scale may be enough
to observe clear local signature of {\em both} magnetic and pairing
correlations. Tuning experiments to be in the regime where the onsite
interactions are of the order the bandwidth ($U \sim 8 t$) provides
the sharpest signal of the many-body effects. 
Our computation of the
entropy $S$ indicates that adiabatic cooling occurs in the
2D Hubbard Hamiltonian with a position dependent
(confining) potential as the interaction strength $U$
is increased via, {\it e.g.}, a Feshbach resonance.  This allows our conclusions
concerning the observability of spin and pairing order to be relevant
 to experiments at constant entropy.

\begin{acknowledgments}
This work was supported under ARO Award W911NF0710576 with funds from
the DARPA OLE Program, by the DOE SciDAC program under grant
DOE-DE-FC0206ER25793, by the US Office of Naval Research,
and by the National Science Foundation under Grant No.\ OCI-0904597 
and through TeraGrid resources provided by NICS under grant number
TG-DMR100007. M.R. and R.T.S. thank the Aspen Center for Physics for
hospitality. C.N.V. acknowledges NSF dissertation enhancement grant
No.\ 0803230.
\end{acknowledgments}

\end{document}


\title{Supplementary material for EPAPS\\ Magnetism and pairing of
  two-dimensional trapped fermions}

\author{Simone Chiesa}
\affiliation{Department of Physics and Astronomy, University of
  Tennessee, Knoxville, TN 37996, USA}
\affiliation{Department of Physics, University of California, Davis,
  CA 95616, USA} 

\author{Christopher N. Varney}
\affiliation{Department of Physics, Georgetown University, Washington,
  DC 20057, USA} 
\affiliation{Joint Quantum Institute, University of Maryland, College
  Park, MD 20742, USA} 

\author{Marcos Rigol}
\affiliation{Department of Physics, Georgetown University, Washington,
  DC 20057, USA} 

\author{Richard T. Scalettar}
\affiliation{Department of Physics, University of California, Davis,
  CA 95616, USA} 

\maketitle

\paragraph*{Shortest non-trivial correlations}
If it would be possible to address low enough temperatures that the
various correlations in the trap would all saturate to their largest
value, we would then use an order parameter to establish where and how
the system is most likely to order.  Experiments are far from that
regime and we therefore focus on the task on selecting short-range
correlations which are representative (in the sense of having the
right spatial, temperature and $U$ dependence) of the behavior of a
true order parameter.

\begin{figure}[!b]
 \centering
 \includegraphics[width=7.5cm,angle=0]{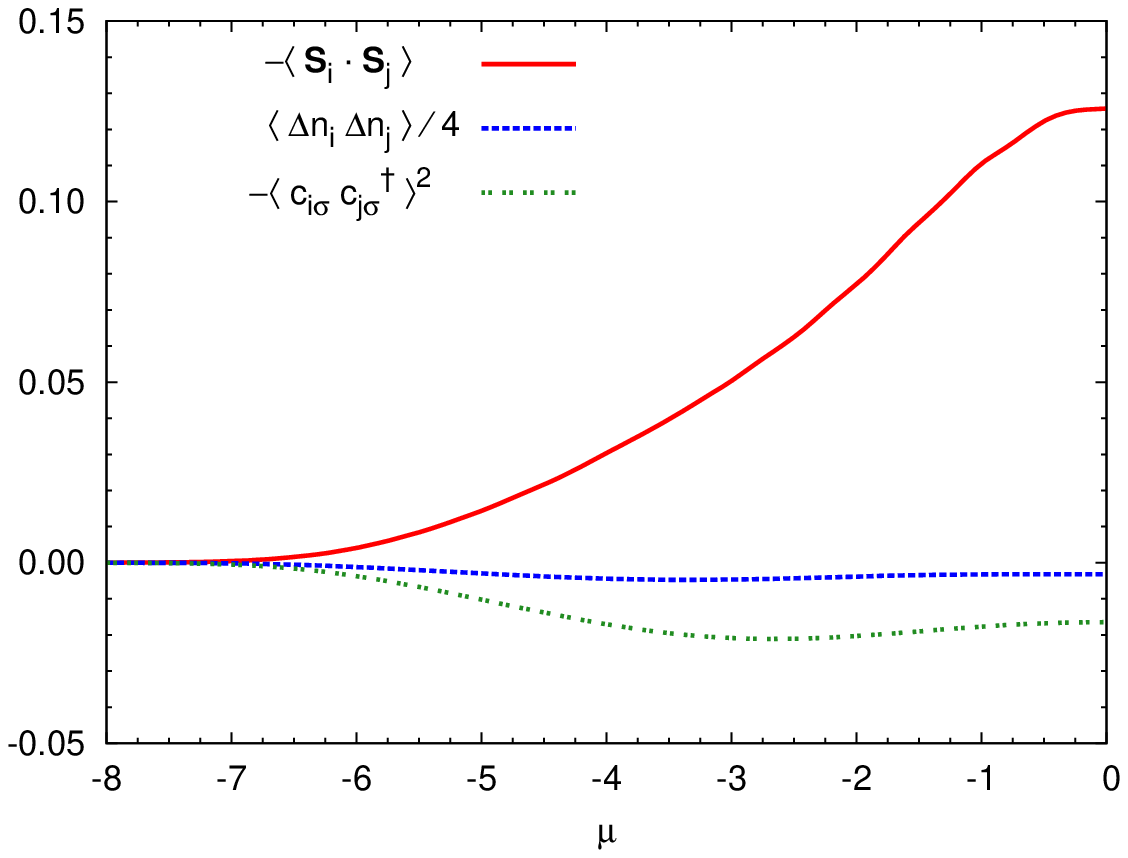}
\caption{\label{FigS1}
  The three contributions to the connected pairing correlation
  function that corresponds to creating and destroying a pair of
  particles on the same (nearest-neighbor) bond as a function of
  chemical potential for a homogeneous system with $U=6\, t$ and
  $T=0.5\, t$. $\mu=0$ corresponds to half-filling. The large
  contribution from $\langle\mathbf{S}_i \cdot \mathbf{S}_j\rangle$
  suggests that this contribution will dominate the shortest range
  pair correlations and hence that the dependence on temperature and
  $U$ in a trapped system of somewhat longer range $P_{xy}(i)$ should
  be studied to get the clearest insight into incipient pair order.}
\end{figure}

A useful analogy is provided by considering the minimal length scale
at which the spin correlation function reflects the exchange energy
$J=4t^2/U$.  Clearly the temperature dependence of the shortest
separation spin correlator, $C_{00}$ (the local moment) is completely
controlled by the energy scale $U$ rather than $J$, as can be seen by
rewriting it in terms of density properties,
$C_{00}=\braket{n}-2\braket{n_\uparrow n_\downarrow}$.

To develop the same sort of analysis in the case of pairing
correlations, consider now the contribution that comes from creating
and destroying a pair on the same bond. One can derive the following
identity,
\begin{align}
  \langle \Delta_{ij} \Delta^\dagger_{ij}\rangle_c=
  -\braket{\mathbf{S}_i\cdot\mathbf{S}_j}+\frac{1}{4}\langle\Delta n_j
  \Delta n_i\rangle -\frac{1}{2}\sum_\sigma
  \braket{c_{i\sigma}c^\dagger_{j\sigma}}^2 .
\label{PijPij}
\end{align}
The three contributions to $\langle \Delta_{ij}
\Delta^\dagger_{ij}\rangle_c$ are reported in Fig.~\ref{FigS1} as a
function of chemical potential for a homogeneous $8\times 8$-site
cluster, $U=6\, t$ and $T=0.5\, t$.  One can see that the largest
contribution is given by ${\mathbf S}_i \cdot {\mathbf S}_j$,
especially in the most interesting density regimes at and near
half-filling. This simple analysis reveals that not only is the
energy scale trivially inherited from that of spin fluctuations but
also that the location of the maximum is pinned where ${\mathbf
  S}_i\cdot {\mathbf S}_j$ is bigger, ${i.e.}$ at half-filling.

\begin{figure}[b]
 \includegraphics[width=3.4cm,angle=0]{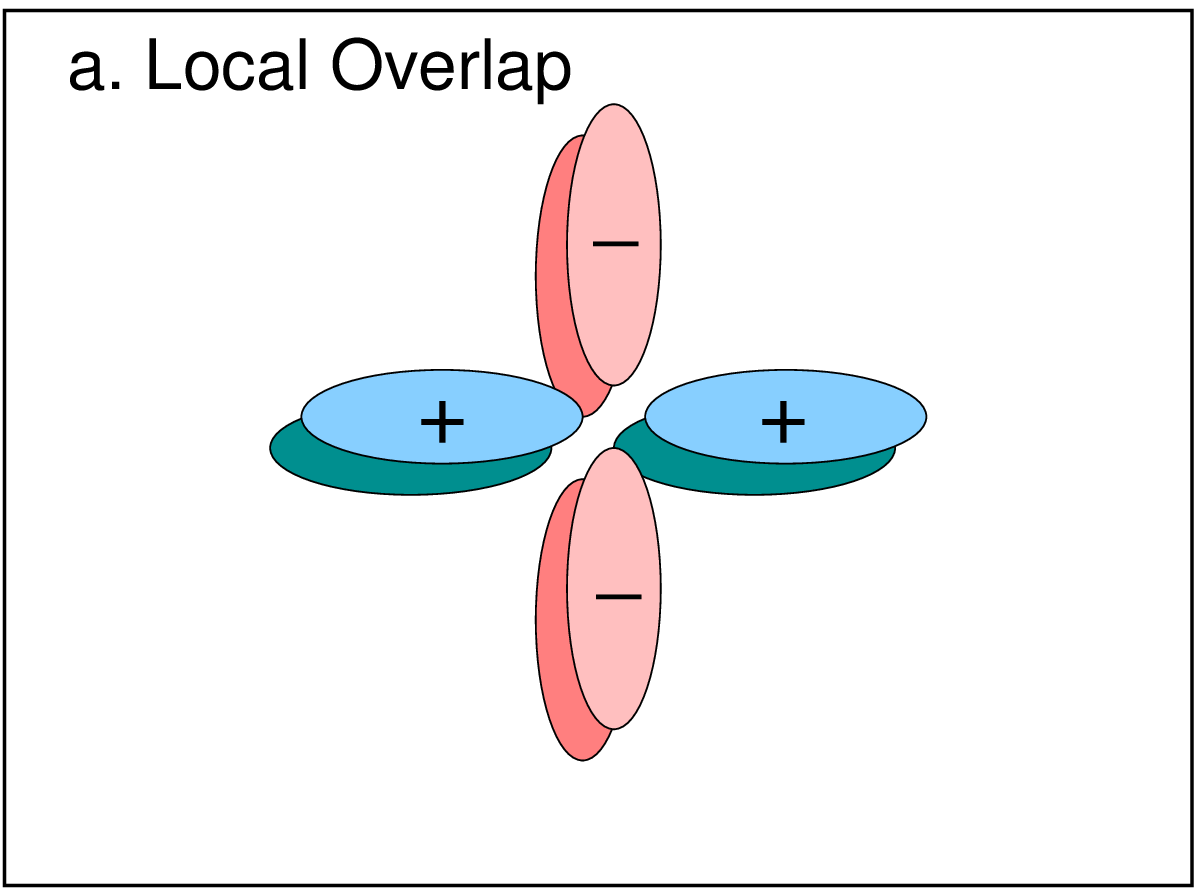}
 \centering
 \includegraphics[width=3.4cm,angle=0]{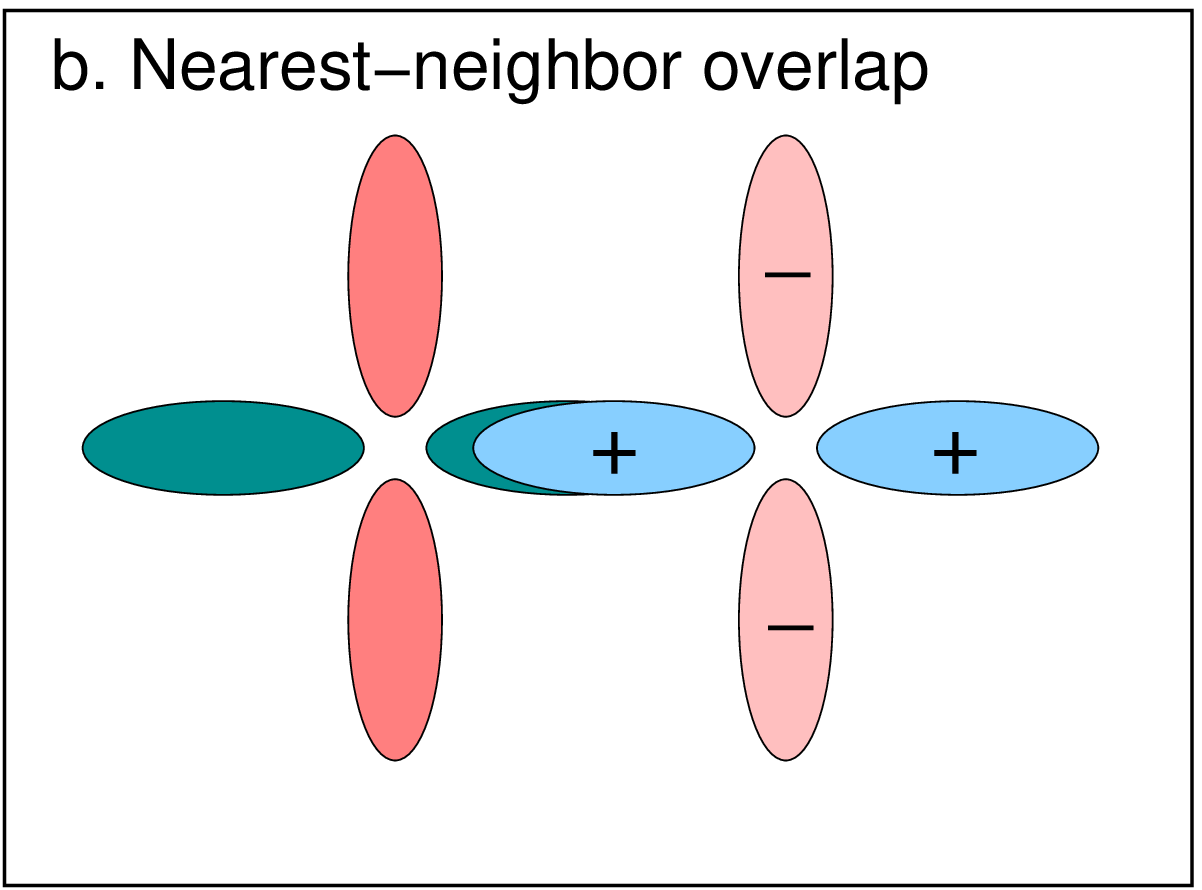}
 \includegraphics[width=8.5cm,angle=0]{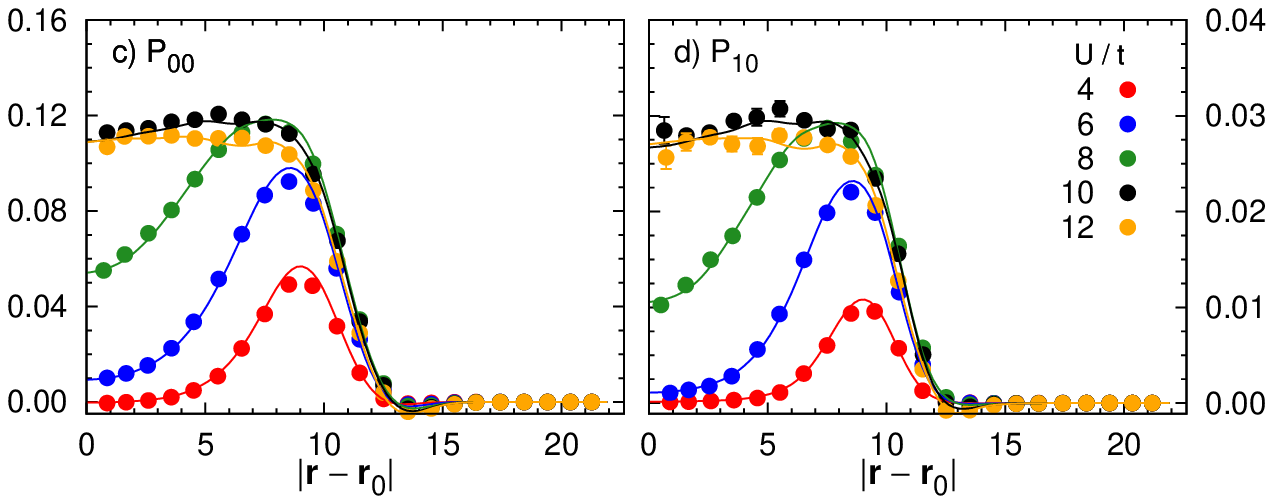}
\caption{\label{FigS2}
  Panels (a) and (b) report a schematic representation of the local
  and nearest-neighbor pair overlap defining $P_{00}(i)$ and
  $P_{10}(i)$.  DQMC values for these quantities are plotted in panels
  (c) and (d).  Their shape as a function of position is very similar
  and strongly reminiscent of $C_{10}(i)$. The drop in amplitude
  perfectly correlates to the decrease in the number of overlapping
  bonds in panels (a) and (b).  As previously suggested by
  Fig.~\ref{FigS1}, both quantities are strongly dominated by
  $C_{10}(i)$ making it difficult to make use of them as a measure of
  the strength of pairing correlations.  }
\end{figure}

\begin{figure}[t]
 \centering
 \includegraphics[width=0.45\textwidth,angle=0]{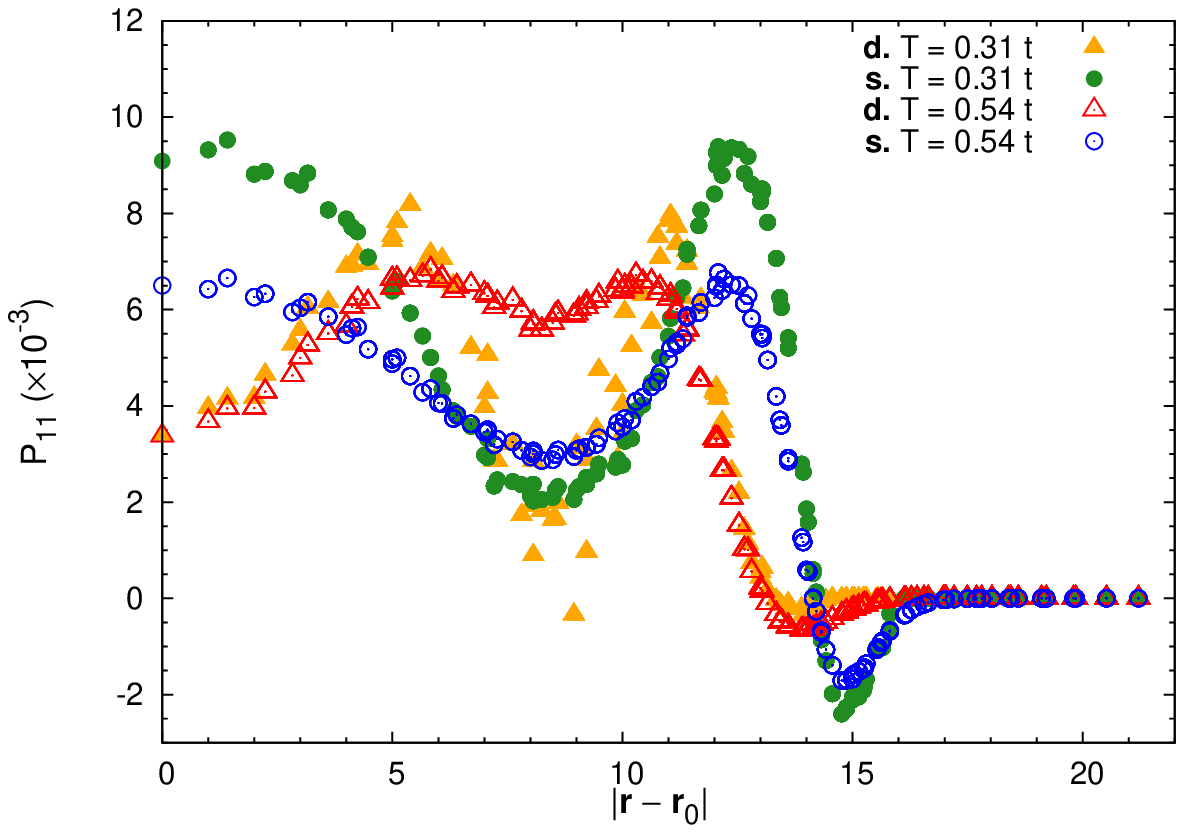}
\caption{\label{Figdwsw}
  Solid symbols: pairing amplitude in the $d$ (orange/triangle) and
  extended-$s$ (green/circle) channels at $T=0.31t$.  Empty symbols:
  pairing amplitude in the $d$ (red/triangle) and extended-$s$
  (blue/circle) channels at $T=0.54t$. Different symmetry of the order
  parameter are peaked in different regions.}
\end{figure}

When using the definition of the spatially dependent $d$-wave pairing
function, the local and nearest-neighbor contributions can be
graphically pictured as in panels (a) and (b) of Fig.~\ref{FigS2}, and
they therefore respectively receive four and one contributions of the
pathological type written in Eq.~(\ref{PijPij}). These pairing
profiles are given in panels (c) and (d) of Fig.~\ref{FigS2}: their
shape mimics the $C_{10}(i)$ profile and the intensity decreases by a
factor of four, reflecting the ratio of overlapping bonds in the upper
panels.  It is thus clear that such correlations cannot be used as a
meaningful measure of pairing in the system and that the shortest
non-trivial correlations are $C_{10}(i)$ and $P_{11}(i)$.

\paragraph*{Extended $s$-wave pairing}
The extended $s$-wave pair creation operator is defined by replacing
$\Delta^d_i$ in the main text by:
\begin{align}
  {\Delta^s_i}^\dagger &=\frac{1}{2}(\Delta^\dagger_{i,i+\hat{x}}
  +\Delta^\dagger_{i,i-\hat{x}}+\Delta^\dagger_{i,i+\hat{y}}+\Delta^\dagger_{i,i-\hat{y}}).
\end{align}

Fig.~\ref{Figdwsw} reports the spatial variation of the $(1,1)$
correlation for both extended-$s$-wave and $d$-wave type at $T=0.54t$
and $T=0.31t$. While both correlations increase with decreasing
temperature in regions surrounding the Mott insulating plateaux (whose
radius is about 8 lattice spacings) the $d$-wave is peaked in narrow
regions centered at density $n=1.0 \pm 0.2$ while the $s$-wave in
broader ones around $n=1.0\pm0.4$. This is in agreement with existing
ground state calculations on related homogeneous models in small
clusters where the extended $s$-wave order parameter was found to be
larger than the one of $d$-wave order in the overdoped regime
\cite{rigol09,dagotto94}. Our present finite temperature analysis
reveals that if an instability were to occur, it would start
developing around $n=0.8$ for $d$-wave pairing and around $n=0.6$ for
extended $s$-wave pairing.

\paragraph*{Calculation of the entropy}
The free energy of a non-interacting system can be computed using the
identity
\begin{align}
  F(\alpha,T)-F(0,T)=\int_0^\alpha \frac{dF(\lambda,T)}{d\lambda}
  d\lambda ,
\end{align}
where $F(\lambda,T)$ is the free energy associated to the hamiltonian
$H(\lambda) = H_0 + \lambda H_1$.

For the Hubbard model we can set $\lambda=U$ and
$H_1 = \sum_i n_{i\uparrow}n_{i\downarrow}$
so that the free energy can be expressed as an integral
of the double occupancy
\begin{align}
  F(\widetilde{U},T)-F(0,T) = \int_0^{\widetilde{U}} dU\; \sum_i \langle
  n_{i\uparrow}n_{i\downarrow}\rangle_{U,T}.
\end{align}

In this paper we have used the LDA to estimate the value of the local
double occupancies since ab-initio DQMC results at the same
temperature have confirmed that such a quantity can be reliably
described in this approximation.

We determine the entropy using the thermodynamic identity
$S=\beta(E-F)$ with $E$, once again, obtained within LDA. Such an
estimate of $E$ is certainly of inferior accuracy if compared to
$\langle n_\uparrow n_\downarrow\rangle$, due to stronger sensitivity
of the kinetic energy on boundary conditions.  However, for those $U$
values for which we have carried out full DQMC simulations in a trap,
we found the accuracy in estimating $E$ to be in the 5\% range and
completely adequate for the kind of conclusions we are drawing in the
main body of this work.